\documentclass{aa}
\usepackage{times}
\usepackage{graphics}

\makeatletter
\let\@internalcite\cite
\def\cite{\@ifstar{\citey}{\citefull}}
\def\citefull{\def\astroncite##1##2{##1\ ##2}\@internalcite}
\def\citey{\def\astroncite##1##2{##1\ (##2)}\@internalcite}
\def\citename{\def\astroncite##1##2{##1}\@internalcite}
\def\@citex[#1]#2{\if@filesw\immediate\write\@auxout{\string\citation{#2}}\fi
  \def\@citea{}\@cite{\@for\@citeb:=#2\do
    {\@citea\def\@citea{; }\@ifundefined 
       {b@\@citeb}{{\bf ??}\@warning
       {Citation `\@citeb' on page \thepage \space undefined}}%
{\csname b@\@citeb\endcsname}}}{#1}}

\makeatother

\begin{document}

\thesaurus{06(13.25.5; 08.02.1; 02.01.2; 08.09.2 Cyg~X-1; 03.13.6)}

\title{Analyzing short-term X-ray variability of Cygnus~X-1 \\
with Linear State Space Models}

\authorrunning{K. Pottschmidt et al.}
\titlerunning{Short-Term Variability of Cygnus~X-1}

\author{Katja Pottschmidt \and Michael K\"onig \and J\"orn Wilms \and
R\"udiger Staubert} 

\institute{Institut f\"ur Astronomie und Astrophysik,
Astronomie, University of T\"ubingen, Waldh\"auser Str.~64, D-72076
T\"ubingen, Germany}

\offprints{K. Pottschmidt,\\ E-mail: katja@astro.uni-tuebingen.de}

\date{Received 7 October 1997 / Accepted 13 February 1998}

\maketitle

\begin{abstract}
  Cyg~X-1 exhibits irregular X-ray variability on all measured timescales. The
  usually applied shot noise models describe the typical short-term
  behavior of this source by superposition of randomly occuring shots 
  with a distribution of shot durations. We have reanalyzed 
  EXOSAT ME observations of Cyg~X-1\ using the more general Linear State
  Space Models (LSSMs). These models, which explicitly take the
  observation-noise into account, model the intrinsic system variability
  with an autoregressive (AR) process. Our fits show that an AR process of
  first order can reproduce the system variability of Cyg~X-1. A possible
  interpretation is again the superposition of individual shots, but with a
  single relaxation time $\tau$. This parameter was found to be
  0.19\,s.         
 

\end{abstract}
    
\section {Introduction}\label{kapitel1}
 
  The X-ray source Cyg~X-1\ was discovered by a rocket flight in 1964
  (\cite{bowyer:65}). In March 1971 the sudden appearance of an accurately
  locateable radio source coinciding with a change in the luminosity of the
  X-ray source led to its identification with the O9.7~Iab supergiant
  HDE~226868 (\cite{hjellming:71,bolton:72}). This star is known as
  part of a single lined spectroscopic binary system with an orbital period
  of 5.6 days at a distance of about 2.5\,kpc. The X-ray emission of Cyg~X-1\
  is produced by the accretion of material from the supergiant primary onto
  the compact object. According to \citey{herrero:95}, the most probable
  masses of the binary partners are 18\,M$_\odot$ for the primary and
  10\,M$_\odot$ for the compact object, which is one of the most
  firmly established black hole candidates (BHCs). The accretion process is
  believed to be fueled by a focused stellar wind from HDE~226868
  (\cite{friend:82}).     

  During most of the time Cyg~X-1 is emitting a non-thermal or hard state
  spectrum (for the soft state, cf.\ \cite{cui:97a}), which can be
  described by a power-law with a photon-index $\Gamma\approx$1.5-1.7,
  modified by an exponential cutoff with a folding energy of about 150\,keV
  and reprocessing features. This spectral form can be interpreted as being
  due to an accretion disk corona (ADC) (\cite{dove:97b} and references
  therein).

  The short-term variability of the X-ray emission has been studied in order to
  gain better insight into the physical processes at work near the
  compact object. Cyg~X-1\ was the first source for which X-ray variability
  on timescales $<$1\,s was detected (\cite{oda:71}). For a long time this
  variability was suspected to be a special black hole signature, but today
  we know that X-ray binaries containing a neutron star instead of a black
  hole can display similar behavior (\cite{stella:85}). Nevertheless, the
  efforts in trying to identify a BHC by its irregular short-term
  variability are continuing (\cite{klis:95}).    
  
  Until now, no special process could be determined that describes all the
  properties of the observed short-term variability. We define
  ``variability on short timescales'' as variability faster than a few
  100\,s, with emphasis on timescales $<$1\,s. These rapid fluctuations are
  usually described in terms of shot noise models. It has become clear,
  however, that in the framework of applying conventional shot noise
  models, complex shot profiles or distributions of shot durations and
  amplitudes have to be assumed to explain the observations
  (\cite{nolan:81,lochner:91,negoro:94}). Shot noise processes exhibit a
  fixed dynamical behavior in the sense that if the shots are given, no new
  dynamical information is produced during the run of the Poisson
  distributed point process. Only by the summation of individual shots
  the temporal correlations in the time series of the process
  are generated (Sect.~\ref{kapitel2.3}).

  We use the alternative method of applying Linear State Space Models
  (LSSMs) which are based on stochastic processes (i.e. autoregressive (AR)
  processes, \cite{scargle:81}) to describe the time series variability. In
  this case the dynamics of the process are produced using a different
  approach: each value of the time series refers to earlier values, with
  their temporal correlation being determined by the dynamical parameters
  of the system (Sect.~\ref{kapitel3.1}).

  The time series of a standard shot noise (exponential decay with one
  relaxation timescale) and an AR process of first order might look very
  similar, although the underlying processes differ. The similarity in the
  time domain leads to shot noise and AR processes both exhibiting an
  exponentially decaying autocorrelation function (\cite{koenig:97a}). In
  contrast to the similarity of the time series, however, the theoretical
  frequency spectra are different (cf.\ Eq.~\ref{pshn} and
  \ref{plss}). Whereas shot noise profiles with varying relaxation
  timescales have to be added to approximate the observed periodogram
  (\cite{lochner:91}), only one dynamical AR parameter is needed to
  reproduce its properties (Sect.~\ref{kapitel3.3} and
  ~\ref{kapitel4.1}).

  We have studied the high time resolution EXOSAT ME lightcurves of
  Cyg~X-1. The data are presented in Sect.~\ref{kapitel2.1}. A commonly
  used method to analyze X-ray variability is to work in the frequency
  domain by fitting theoretical power spectra to the periodogram of a
  lightcurve: The periodogram of Cyg~X-1\ and the power spectra of the
  usually applied shot noise models are reviewed in
  Sect.~\ref{kapitel2.2} and \ref{kapitel2.3}, respectively. Our analysis
  of the lightcurves with LSSMs, however, is not based on fitting the
  periodogram but on an alternative procedure working in the time
  domain. Sect.~\ref{kapitel3} deals with the theory of the LSSM
  analysis. Our results are presented in Sect.~\ref{kapitel4}, and in
  Sect.~\ref{kapitel5} a possible explanation for the derived relaxation
  timescale is discussed combining the temporal LSSM results and
  simulations of comptonized radiation using an accretion disk corona model
  for Cyg~X-1.
   
\section {The short-term X-ray variability of Cyg~X-1}\label{kapitel2}

\subsection{The data}\label{kapitel2.1}

  The EXOSAT raw data have been stored on \emph{Final Observation
  Tapes} (FOTs) and are now available at the \emph{HEASARC}
  archive. Table~\ref{tab1} lists the FOT observations of Cyg~X-1\ that we
  have analyzed with Linear State Space Models. We have chosen the ME
  data\-streams provided by the primary timing telemetry modes HTR3 und
  HER6. These observations only contain events registered in the Argon
  counters (1--20\,keV). The lightcurves are given as countrates normalized
  to one half of the detector array (i.e.\ four Argon counters). Using the
  \emph{Interactive Analysis} (IA) software, we extracted lightcurves
  corrected for dead-time effects and collimator efficiency (for an overview
  of the IA see \cite{parmar:95}). For the purpose of this paper an
  explicit background subtraction is not necessary since the LSSM is
  implicitly modeling the measurement process (see Sect.~\ref{kapitel3.2},
  Eq.~\ref{lssm2}).

  \begin{table}
  \caption{EXOSAT ME observations that have been analyzed with
  LSSMs.}\label{tab1}   
  \begin{tabular}{cccc} \hline
  No. & FOT tape $^{\rm a}$ & \multicolumn{1}{c}{start time (UT)} &
  original time \\
      &  & \multicolumn{1}{c}{[dd.mm.yy hh:mm]} & resolution [ms] \\
  \hline 
  1 & EE8087(05)\makebox[0cm]{\,${}^\star$} & 28.07.83 22:51 & 07.813\\
  2 & EE1723(06)\makebox[0cm]{\,${}^\star$} & 25.09.83 05:15 & 01.953\\   
  3 & EE2461(06) & 21.05.84 04:59 & 07.813\\
  4 & EE2956(06) & 07.07.84 10:57 & 15.630\\
  5 & EE2956(07) & 07.07.84 13:23 & 15.630\\
  6 & EE2929(10) & 09.07.84 14:59 & 11.720\\
  7 & EX3620(08)\makebox[0cm]{\,${}^\star$} & 24.07.84 20:20 & 07.813\\
  8 & EE3962(13) & 02.11.84 19:20 & 09.766\\ 
  9 & EE3962(16) & 02.11.84 20:27 & 11.720\\
  \hline
  \end{tabular}

  $^{\rm a}$ Final observation tape number and number of observation
  therein\\
  ($^\star$: telemetry mode HTR3, else: HER6). 
  \end{table} 

\begin{figure} \resizebox{\hsize}{!}{\includegraphics{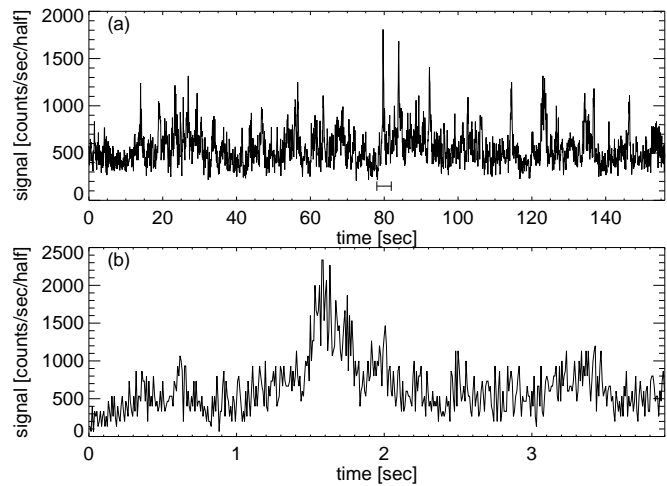}}
  \caption{\textbf{a} X-ray lightcurve of Cyg~X-1, showing a part of the
  fifth observation on FOT Nr.~EE8087 (No.~1 of Table~\ref{tab1}, energy
  range=1--20\,keV, bin time=62.4\,ms, length=156\,s). \textbf{b} Blow-up
  of the marked segment of subfigure a with a length of 3.9\,s and the
  original bin time of 7.8\,ms, showing a typical shot structure. Both
  lightcurves are plotted without error bars for reasons of
  clarity.}\label{fig1} \end{figure}

  All EXOSAT observations of Cyg~X-1 have found the source in its
  usual hard state. An example for the characteristic hard state variability of
  Cyg~X-1\ on short timescales can be seen in Fig.~\ref{fig1}.
  
\subsection{The periodogram of Cyg~X-1}\label{kapitel2.2} 

  The periodogram $P(f_k)$ of a lightcurve $y(t_j)$ (where $j=1, \ldots,N$)
  is providing the ``strength'' of harmonic variations with a certain 
  frequency $f_k$ in the lightcurve. $P(f_k)$ is defined as the squared
  modulus of the discrete Fourier transform of $(y(t_j)-\bar{y})$ and is
  calculated for the Fourier frequencies $f_k=k/(N\cdot\Delta t)$ with bin
  time $\Delta t$ and $k\in\{1,\ldots,M\}$, where $M$ is the integer part
  of $N/2$ (\cite{scargle:82}):    
  
  \begin{equation}\label{per}
  P(f_k)=\frac{1}{N}\left|\sum_{j=1}^N{(y(t_j)-\bar{y})e^{2\pi{i}f_kt_j}}\right|^2 
  \end{equation}

\begin{figure}
  \resizebox{\hsize}{!}{\includegraphics{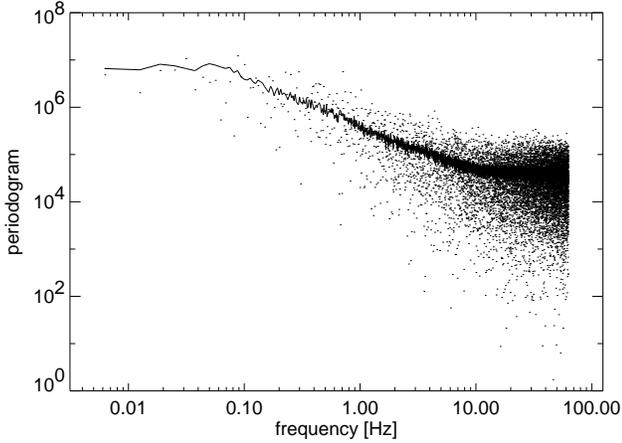}}
  \caption{Sample periodogram (solid line) of Cyg~X-1, averaged over 48
  individual periodograms, which have been obtained by dividing the
  original lightcurve (No.~1, Table~\ref{tab1}) into 48 segments of 156\,s
  each and calculating $P(f_k)$ for each segment (Eq.~\ref{per}). The
  values of a typical individual periodogram, fluctuating wildly, are also
  displayed (dots).}\label{fig2}    
  \end{figure}

  Fig.~\ref{fig2} shows the logarithmically plotted sample periodogram
  (solid line) of the observation No.~1 (Table~\ref{tab1}). It has been
  obtained by averaging over periodograms $P(f_k)$ from 48 different
  lightcurve segments (details see caption of Fig.~\ref{fig2}). Since
  $P(f_k)$ is a $\chi_2^2$-distributed random variable, its standard
  deviation is equal to its mean (\cite{klis:89a}). Therefore, an
  individual periodogram exhibits large fluctuations (Fig.~\ref{fig2},
  dots). Calculating the sample periodogram significantly reduces the
  scatter and allows the possibility to estimate the theoretical spectrum of
  the process responsible for the observed variability. 

  The periodogram of the short-term variability of Cyg~X-1\ in its hard
  state is well known and exhibits the following distribution of timescales
  (Fig.~\ref{fig2}): \begin{itemize} \item For frequencies above 10\,Hz
  white noise dominates the periodogram which means that the variability of
  the lightcurves on all corresponding timescales is almost equally
  strong. These fluctuations (i.e.\ photon statistics, particle background)
  are introduced by the measurement procedure. It is common practice to
  subtract a constant (corrected for deadtime effects) which represents
  this noise component from the periodogram (\cite{belloni:90a,zhangw:95a}).
  \item Towards lower frequencies the power of the variability increases
  within the frequency range of roughly 0.04--10\,Hz. This behavior is
  called red noise. It is often modeled by a power law,
  $f^{-\alpha}$. Variability of the $f^{-\alpha}$-type is known from many
  X-ray binaries (\cite{klis:95}) and also from active galactic nuclei
  (\cite{mchardy:89}). In the case of Cyg~X-1, $\alpha$ is approximately 1
  (\cite{nolan:81,lochner:91}).
  \item An important feature indicating the stationarity of the short-term
  variability process is the flattening of the power spectrum for
  frequencies below $\approx$0.04\,Hz. This flat top corresponds to the
  absence of additional long-term variations in the lightcurves. There are,
  however, some EXOSAT ME observations of Cyg~X-1\ which show low
  frequency noise in the form of increasing power towards lower frequencies
  below 0.001\,Hz (\cite{angelini:94}). Since this component is not always
  present, it is most likely not produced in the same physical process as
  the short-term variability. Suggested causes for this low frequency
  variabilities are instabilities in the mass transfer process. Sometimes
  the low frequency noise is associated with absorption dips in the
  lightcurve (\cite{kitamoto:84,angelini:94}).
  \item Several authors reported a transient quasi-periodic oscillation
  (QPO) feature in the power spectrum with a central frequency of about
  0.04\,Hz (\cite{kouveliotou:92,angelini:94,vikhlinin:94,borkus:95}),
  whereas other authors have not found any evidence for QPOs
  (\cite{belloni:90a,miyamoto:92}). Thus the significance of this feature
  is not yet clear. In Fig.~\ref{fig2} no QPO feature is present.
  \end{itemize}
  
\subsection{The shot noise model}\label{kapitel2.3} 
 
  \citey{terrell:72} proposed that the structures observed in the lightcurves
  of Cyg~X-1 might be due to a shot noise process $s(t)$, i.e.\ the
  superposition of randomly occuring shots with the shot profile $h(t')$: 
  \begin{equation}\label{shn}
  s(t)=\sum_ih(t-t_{\rm SH})
  \end{equation}
  Here the $t_{\rm SH}$ are the times at which the shots occur, with the time
  intervals between the $t_{\rm SH}$ following a Poisson distribution. 

  The standard shot profile $h(t')_{\rm ST}$ is identical for all shots and
  consists of an instantaneous rise to height $h_0$ and an exponential
  decay with the decay time $\tau$:
  \begin{eqnarray}
  h(t')_{\rm ST} & = & \left\{ \begin{array}{ll}
        0 & , t'<0\\
        h_0 \cdot e^{-t'/\tau} & , t'\geq 0
  \end{array} \right.\label{shp}    
  \end{eqnarray}
 
  \begin{figure} \resizebox{\hsize}{!}{\includegraphics{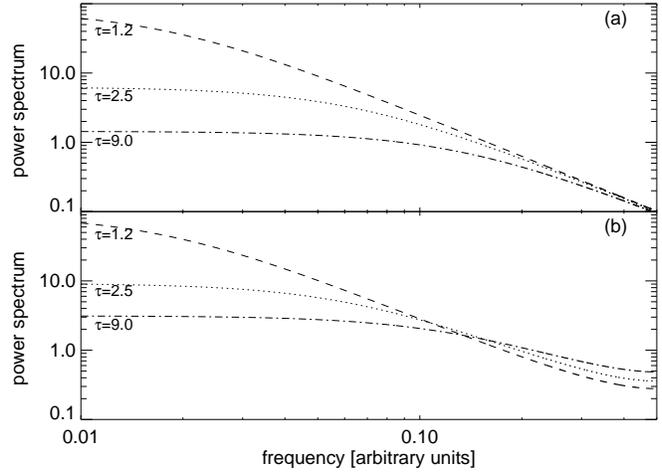}}
  \caption{\textbf{a} Shot noise power spectra for three different values of
  $\tau$, calculated according to Eq.~\ref{pshn} ($\lambda$=1,
  $h_0$=1). \textbf{b} LSSM power spectra for the same values of $\tau$ as in
  (a), calculated according to Eq.~\ref{plss} ($\sigma_\epsilon^2$=1). All
  power spectra are given in arbitrary units and $\Delta t$ has been
  normalized to 1. In subfigure b, the variance of the observation-noise,
  $\sigma_\eta^2$, has been set to 0 to allow direct comparison of a and
  b.}\label{fig3} \end{figure} 

  Shot noise models are often evaluated by comparing their theoretical power
  spectrum to the observed power spectrum
  (\cite{belloni:90a,lochner:91}). The theoretical power
  spectrum of the standard shot noise process is (\cite{lehto:89}):  
  \begin{equation}\label{pshn}
  S(f)_{\rm SN}\propto\frac{\lambda{h_0}^2}{(1/\tau)^2+(2\pi f)^2}, \quad
  f\neq0, 
  \end{equation}
  where $1/\lambda$ is the average time interval between the shots.
  $S(f)_{\rm SN}$ provides the flat top for $f\ll1/(2\pi \tau)$ but has a
  fixed logarithmic slope of $\alpha=$2 for higher frequencies
  (Fig.~\ref{fig3}a). Since the shot noise process $s(t)$ is continuous
  in $t$, whereas the observed lightcurve is the result of integrating the
  measured counts over a finite number of intervals $\Delta t$,
  $S(\omega)_{\rm SN}$ must be corrected for binning. \citey{lochner:91}
  have shown, that binning makes the shot noise power spectrum even steeper
  for high frequencies.  

  Neither the observed slope nor the white noise floor of the Cyg~X-1
  periodogram (Fig.~\ref{fig2}) can be reproduced by the standard shot
  noise process. In frequency domain fits the white noise level is usually
  treated as an additional constant parameter (cf.
  Sect.~\ref{kapitel2.2}). To model the observed slope, different shot
  profiles with distributions of shot durations and shot amplitudes have
  been proposed. A number of those models, each having many degrees of
  freedom, are able to reproduce the observed periodogram and other second
  order statistics (\cite{belloni:90a}, \cite{lochner:91}). Higher order
  statistics like the time skewness (\cite{priedhorsky:79}) or phase portraits
  (\cite{lochner:91}) were also studied but no special shot noise model
  could be singled out that allows for a homogeneous description of
  different observations. In Sect.~\ref{kapitel3.3} we discuss the
  theoretical power spectrum of a first order LSSM and show that it can
  reproduce all the features of the hard state periodogram of Cyg~X-1,
  requiring only one temporal parameter $\tau$ (Fig.~\ref{fig3}b).

\section {The Linear State Space Model (LSSM)}\label{kapitel3}
  
  The LSSM analysis is a recently developed tool to model stochastic
  time series (\cite{gantert:93}). It has e.g.\ been used to describe
  medical time series (hand-tremor data,\cite{timmer:95b}). 
  \citey{koenig:97a} were the first to apply the new method to astronomical
  data by successfully fitting lightcurves of active galactic nuclei with
  first order LSSMs (see also \cite{koenig:97c}). The mathematical
  background of the LSSMs and the associated fitting-procedure have been
  discussed in detail by \citey{koenig:97a} and in the references
  therein. Here, we shall concentrate on first order LSSMs, for we found
  them to be appropriate to describe the Cyg~X-1\ lightcurves (see
  Sect.~\ref{kapitel4}).   
  
\subsection{The autoregressive process}\label{kapitel3.1}
 
  One possibility to model irregular variability (i.e.\ temporal variations
  that cannot be forecasted exactly) is to assume that it is caused by a
  chaotic process (\cite{voges:87,unno:90}). The LSSMs, on the other
  hand, are based on the alternative assumption, that the variability of
  the observed system is produced by a stochastic process. The LSSMs use the
  rather general autoregressive (AR) formulation for the stochastic
  system-light\-curve $x(t_j)$.  

  The AR processes were first introduced by \citey{yule:27} to model the
  variability of Wolf's sunspot numbers and have been well studied since
  then (e.g.\ by ~\cite{scargle:81}). An \emph{AR process of order $p$}
  (AR[p]) is defined by:  

  \begin{equation}\label{ar}
  x(t_j)=\sum^p_{i=1}a_i \cdot x(t_j-i\cdot\Delta t) + \epsilon(t_j),
  \hspace{0.1cm}\epsilon(t)\propto {\cal N}(0,\sigma_\epsilon^2)
  \end{equation}
  The time series $x(t_j)$ is sampled at discrete times $t_j$ with time
  resolution $\Delta t$. The purely stochastic component $\epsilon(t)$, is
  a Gaussian random variable with mean 0 and variance
  $\sigma_\epsilon^2$. Since Eq.~(\ref{ar}) is of the same structure as
  a regression equation for the variables $x(t_j)$ and $(x(t_j-i\cdot\Delta
  t))$ the name \emph{autoregressive} process has been assigned to it.

  In an AR lightcurve, for each $t_j$ the value of the random variable $x(t_j)$
  is correlated with the values of the process at earlier times. This  
  correlation is decreasing with increasing time differences between two
  lightcurve values (expressed by an exponentially decaying autocorrelation
  function). The stochastic component $\epsilon(t)$ is the reason that the
  process does not simply come to rest. $\epsilon(t)$ is an intrinsic
  property of the system-variability: the system-noise.       
 
  The temporal correlations in the AR lightcurve are characterized by the
  dynamical parameters $(a_i)$. The $p$ dynamical parameters are related to
  $p$ temporal parameters, describing the temporal structures in the
  lightcurves: depending on the process, the $(a_i)$ represent  
  stochastic relaxators with relaxation times $\tau$, or damped stochastic
  oscillators with relaxation times $\tau$ and periods $P$, or both
  (\cite{honer:94}). In case of an first order AR process there
  is only one dynamical parameter $a_1=a$. For stationary processes $|a|$
  must be $<$1 and only positive values of $a$ lead to plausible physical
  models. The corresponding temporal parameter, $\tau=-1/\ln|a|$, is the
  relaxation time of a stochastic relaxator, one representation of which is
  an exponentially decaying shot as described by Eq.~(\ref{shp}). 
             
\subsection{The observation-noise}\label{kapitel3.2}
  
  For the determination of the temporal parameters of the system-process
  the noise which is caused by the measurement (i.e. photon statistic,
  particle background) has to be considered because it disturbs the
  temporal structures of the system-lightcurve. If the observation of a
  system-lightcurve is modeled with a plain AR process, the so called
  observation-noise will lead to an underestimation of the true
  temporal system-parameters (\cite{robinson:79,koenig:97a}). In order to
  solve this general problem, a LSSM consists of two equations: the
  system-equation for the intrinsic system-lightcurve and the
  observation-equation for the measured lightcurve. The latter describes
  the observed lightcurve $y(t_j)$ by \emph{explicitly considering the
  influence of the observation-noise} $\eta(t_j)$ on the system-lightcurve.

  The important model for the analysis of the short-term variability of
  Cyg~X-1\ is the \emph{LSSM of first order}:
  \begin{eqnarray}
  x(t_j) & =& a \cdot x(t_j-\Delta t) + \epsilon(t_j),
  \quad \epsilon(t_j) \propto {\cal N}(0,\sigma_\epsilon^2)\label{lssm1} \\
  y(t_j) & =& c \cdot x(t_j) + \eta(t_j),
  \hspace{1.15cm}\eta(t_j)\propto {\cal N}(0,\sigma_\eta^2)\label{lssm2}
  \end{eqnarray}
  Like the system-noise $\epsilon(t_j)$, the observation-noise $\eta(t_j)$ is
  modeled by a Gaussian noise component; $c$ is a constant normalization
  factor. The equations for higher order LSSMs are given by
  \citey{koenig:97a} (the temporal parameters of an LSSM of order $p$
  (LSSM[p]) correspond to the $p$ dynamical parameters of Eq.~\ref{ar}).

\subsection{The power spectrum}\label{kapitel3.3}

  The power spectrum $S(f)_{{\rm LSSM[1]}}$ of a first order Linear State
  Space Model has the following form (\cite{koenig:97a}):   
  \begin{equation}\label{plss}
  S(f)_{{\rm LSSM[1]}}\propto\frac{\sigma_\epsilon^2}{1+a^2-2\,a\,{\rm
  cos}(2\pi f)}+\sigma_\eta^2 
  \end{equation} 
  The examples for $S(f)_{{\rm LSSM[1]}}$ displayed in Fig.~\ref{fig3}b show
  the flat top as well as red noise. The observational white noise floor is
  also provided by Eq.~\ref{plss} but has been omitted in
  Fig.~\ref{fig3}b ($\sigma_\eta^2=$0) to allow the direct comparison
  with the standard shot noise spectra (Fig.~\ref{fig3}a): In contrary to
  the fixed red noise slope $\alpha=$2 of the standard shot noise spectra,
  the slope of $S(f)_{{\rm LSSM[1]}}$ can be modeled by adjusting $a$,
  i.e. $\tau$. All the features of the hard state power spectrum of Cyg~X-1
  can thus be reproduced by the power spectrum of the LSSM[1] which is
  defined by only one temporal parameter $\tau=-1/\log|a|$. Furthermore,
  Eq.~\ref{plss} does not need to be corrected for binning effects, since the
  definition of the LSSM already is discrete in $t$ (see
  Eq.~\ref{lssm1} and \ref{lssm2}).
 
  Principally, LSSMs can be evaluated by fitting their power spectra to
  the measured periodogram. However, fits in the frequency domain do not
  allow the explicit modeling of the observational noise and may be
  influenced by several other sources of uncertainties, especially by
  spectral leakage, which contaminates the power spectrum of each finite
  time series (\cite{deeter:82,deeter:84}). To avoid these
  uncertainties our LSSM analysis was performed in the time domain by
  direct comparison of the model-lightcurves and the measured
  lightcurves. This method allows fits with a statistical significance
  higher than that of the frequency domain fits (\cite{koenig:97a}).

\subsection{The fitting procedure}

  The parameters of a LSSM[p] fit are estimated with the help of the maximum
  likelihood procedure: the set of parameters is derived, for which the
  probability of observing the measured lightcurve $y(t)$ is at maximum. We
  have obtained those parameters using the expectation-maximization (EM)
  algorithm (\cite{honer:94}). The method is iterative, starting from
  a set of initial parameter values. In the expectation step, the Kalman 
  filter is applied, which allows to estimate the unobservable
  system-lightcurve $x(t_j)$ by using the observed lightcurve $y(t_j)$ and a
  given set of LSSM parameters. In the following maximization step the
  likelihood function is maximized, taking the estimated $x(t_j)$ into
  account and providing a new set of parameters. The results of this
  iterative procedure are best estimates for the LSSM[p] parameters
  and for the intrinsic system-lightcurve (a detailed example is discussed in
  Sect.~\ref{kapitel4.1}). Note, that the latter is \emph{not} produced
  using a smoothing filter: all variability timescales of the observed
  lightcurve (see e.g.\ Fig.~\ref{fig4}a) are still present in the
  derived system-lightcurve (see e.g.\ Fig.~\ref{fig4}b).   
 
  The length and the time resolution of lightcurves, which are analyzed by
  LSSMs, have to be chosen considering the following requirements:
  \begin{itemize} \item The duration of the lightcurve has to be several
  times the relaxation timescale, which is expected to be a few tenths of a
  second (shot noise models), or the relaxator cannot be found.  \item The
  bin time has to be well below the relaxation timescale to allow us to
  distinguish the relaxator from statistical fluctuations.
  \end{itemize} Simulations by \citey{koenig:97b} have shown that the
  duration should be at least 5 times larger and that the bin time should
  be at least 10 times smaller than $\tau$. In order to fulfill these
  criteria without an unreasonable increase in LSSM computing time, we have
  fitted lightcurves with a bin time of 16\,ms and length of 16\,s (e.g.\
  Fig.~\ref{fig4}a). Although this time interval is shorter than the
  intervals typically studied in the context of shot noise analysis, it is
  long enough for the determination of the relaxator.

\section {Results}\label{kapitel4}

\subsection{LSSM fits of various order to one exemplary
  lightcurve}\label{kapitel4.1} 

  \begin{figure} 
  \resizebox{\hsize}{!}{\includegraphics{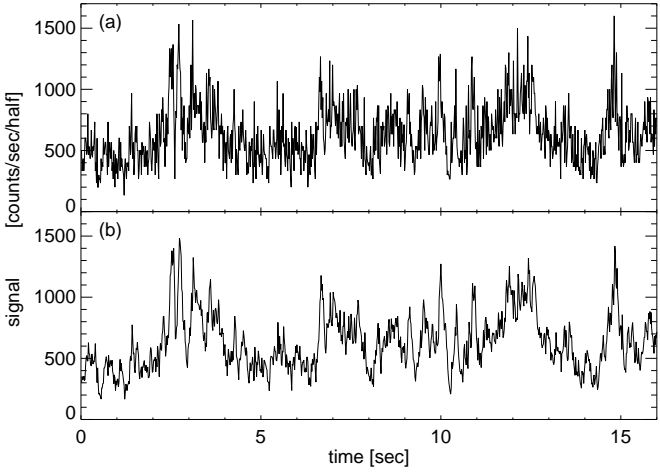}}
  \caption{\textbf{a} One of the 900 EXOSAT ME Cyg~X-1\ lightcurves with
  a length of 16\,s and a bin time of 16\,ms, which were modeled by
  LSSMs. Mean countrate: 633.1\,counts/sec/half, standard deviation:
  234.5\,counts/sec/half. \textbf{b} Best estimate for the autoregressive
  system-lightcurve, which has been obtained by applying a LSSM of order
  $p=$1 to the lightcurve displayed above (see Table~\ref{tab2}). Both
  lightcurves are shown without error bars for reasons of
  clarity.}\label{fig4} 
  \end{figure}
 
  Table~\ref{tab2} shows the results of LSSM fits of order 0--5 to the
  lightcurve segment displayed in Fig.~\ref{fig4}a: The relaxation
  timescale $\tau$ of about 0.2\,s, which is derived by the LSSM[1] fit, is
  also found by the fits of higher order (the LSSM[0] represents
  uncorrelated white noise and therefore does not yield a temporal
  parameter). All other temporal correlations are decaying within a few
  time bins. Since they are due to statistical fluctuations they are
  negligible (\cite{koenig:97b}). This means, that the dynamics of the
  lightcurve can most likely be described by one relaxation timescale. If
  this is the case, the LSSM[1] fit should deliver an adequate description of
  the observed lightcurve and the goodness of fit should not increase
  significantly for higher order LSSMs.  

  \begin{table} 
  \caption{Results of LSSM fits of various order to the lightcurve in
  Fig.~\ref{fig4}a.\label{tab2}}   
  \begin{tabular}{cccc}
  \hline
  \multicolumn{1}{c}{Order $p$ of}& \multicolumn{1}{c}{Relaxator $\tau$}&
  \multicolumn{1}{c}{Period $P$}   & \multicolumn{1}{c}{KS probability}\\  
  \multicolumn{1}{c}{the LSSM}    & \multicolumn{1}{c}{[s]}&
  \multicolumn{1}{c}{[s]}& \multicolumn{1}{c}{$\%$}\\ 
  \hline
       0&                -&           -&    0.0\\
  \hline
       1&            0.185&           -&   93.5\\
  \hline
       2&            0.209&           -&   87.0\\
        &            0.018&           -&       \\
  \hline 
       3&            0.177&           -&   97.6\\
        &            0.022&       0.092&       \\     
  \hline
       4&            0.204&           -&   88.5\\
        &            0.046&       0.128&       \\
        &            0.011&           -&       \\
  \hline
       5&            0.175&           -&   97.4\\
        &            0.046&       0.128&       \\
        &            0.015&       0.096&       \\
  \hline
  \end{tabular}
  \end{table}

  A statistical test on the residuals, i.e.\ the difference between the
  derived system-lightcurve and the observed lightcurve, was performed to
  determine the quality of the LSSM fits: If a given LSSM can describe the
  observation, the residuals should be an uncorrelated white noise
  realization (Eq.~\ref{lssm2}). In this case all the temporal correlations
  of the observed lightcurve are modeled by the system-lightcurve. We use
  the Kolmogorov-Smirnov (KS) test to quantify the goodness of the LSSM
  fits. It compares the power spectrum of the residuals to a flat white
  noise power spectrum.

  In Table~\ref{tab2} we list the KS probabilities for the residual
  lightcurves to actually be white noise realizations. For the
  LSSM[1] fit a very high probablity of 93.5$\%$ is reached, which means
  that the observed lightcurve can be well modelled by an LSSM[1]
  process. This is consistent with the visual impression from
  Fig.~\ref{fig4}: the LSSM[1] estimate (b) tracks the observed lightcurve
  (a) very closely. Higher order LSSM fits do not improve the the fit
  significantly. The LSSM[0] (white noise) is of course rejected at any
  level of confidence.

\begin{figure} \resizebox{\hsize}{!}{\includegraphics{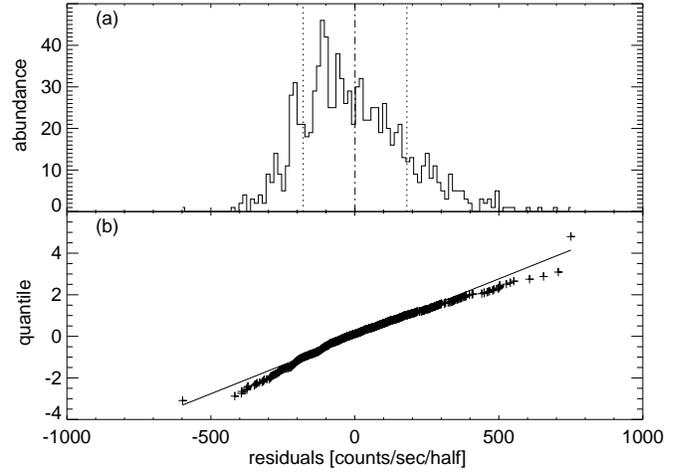}}
  \caption{\textbf{a} Distribution and \textbf{b} normal quantile plot of
  the residuals of the LSSM[1] fit to the lightcurve given in
  Fig.~\ref{fig4}a. The lines in subfigure a indicate the mean and standard
  deviation of the observation-noise. The straight line in subfigure b
  corresponds to a Gaussian distribution having the same mean and variance
  as the distribution of the residual lightcurve.}\label{fig4neu}
  \end{figure}

  The distribution of the LSSM[1] model residuals is shown in
  Fig.~\ref{fig4neu}a. Its mean and variance are 0.7\,counts/sec/half and
  (179.9\,counts/sec/half)$^2$, respectively. Thus the variance of the
  observation-noise is about 82$\%$ of the upper limit as given by the mean
  countrate of the original lightcurve and assuming a pure Poisson
  process. The lower half of Fig.~\ref{fig4neu} displays the normal
  quantile plot of the fit residuals: it arranges the residuals in
  increasing order and for each data value indicates the position that
  corresponds to the same probability in a normal distribution. If the data
  are normally distributed, all points should lie on a straight
  line. Fig.~\ref{fig4neu}b shows that the distribution of the residuals
  can be approximated by a Gaussian distribution, but is compressed for low
  and stretched for high countrates. This indicates that the fit residuals
  are Poisson-distributed as expected for the observation-noise (mean
  countrate of the original lightcurve: 10.1\,counts per 16\,ms
  timebin). The LSSM[1] model parameters reflect the Poisson distribution
  of the observation-noise by underestimating the Gaussian variance
  $\sigma_\eta^2$ given as (143.8\,counts/sec/half)$^2$ in this example.
                                               
\subsection{LSSM fits of first order to a sample of
  lightcurves}\label{kapitel4.2} 
 
  In order to obtain a reliable value for the temporal parameter $\tau$,
  we have fitted LSSM[1]s to a sample of 900 lightcurve segments covering a
  total length of four hours. We have used 1600\,s of each of the nine
  observations listed in Table~\ref{tab1}. The 1600\,s consist of 100
  equally long, uninterrupted segments with a temporal resolution of 16\,ms.

  For each of the nine observations 100 values for the
  relaxator $\tau$ were obtained. In Table~\ref{tab3}, the mean $\bar\tau$,
  the standard deviation $s_\tau$, and the most probable value $\tau_{\rm
  max}$ of these distributions are listed. Averaging $\bar\tau$, $s_\tau$,
  and $\tau_{\rm max}$ over the nine observations gives (0.28$\pm$0.04)\,s,
  (0.13$\pm$0.03)\,s, and (0.19$\pm$0.03)\,s, respectively. All
  distributions exhibit the same asymmetric form, in the sense that
  $\tau_{\rm max}$ is always lower than $\bar\tau$ but still higher than the
  1$\sigma$ deviation. This kind of distribution is consistent with the
  results that have been obtained from simulations of AR
  lightcurves based on one temporal parameter $\tau$ (\cite{hamilton:87}).    

  The lightcurves of three of the observations (No.~1,~2, and 7 of
  Table~\ref{tab3}) take the whole Argon energy range (1--20\,keV)
  into account, i.e. their absolute countrates may be compared. Even though
  the analyzed segments of one of those three observations (No.~7) have a
  considerably smaller mean countrate than those of the other two, the
  corresponding distribution of $\tau$ is not significantly different. This
  means that no correlation between $\tau$ and the luminosity can be seen,
  which is confirmed within each of the nine observations: the mean
  countrates and the relaxation times of the 100 segments are not correlated. 

  \begin{table}
  \caption{Results of the LSSM(AR[1]) fits to EXOSAT ME
  observations of Cyg~X-1.}\label{tab3}
  \begin{tabular}{cccccc} \hline
  No. $^{\rm a}$ & energy channels &
  count- & $\bar\tau$ & $s_\tau$ & $\tau_{\rm
  max}$ \\ 
  & & rate$^{\rm b}$ & [s] & [s] & [s] \\
  \hline 
  1 & 1--128& 565$\pm$57 & 0.26 & 0.10 & 0.22 \\
  2 & 1--128& 567$\pm$37 & 0.36 & 0.17 & 0.22 \\   
  3 & 9--21, 22--51& 541$\pm$33 & 0.30 & 0.16 & 0.22 \\
  4 & 5--19, 27--65& 388$\pm$19 & 0.26 & 0.12 & 0.17 \\
  5 & 5--16, 27--65& 457$\pm$29 & 0.31 & 0.16 & 0.19 \\
  6 & 7--21, 26--81& 499$\pm$35 & 0.25 & 0.11 & 0.17 \\
  7 & 1--128& 389$\pm$24 & 0.25 & 0.09 & 0.22 \\
  8 & 5--16, 17--49& 359$\pm$14 & 0.27 & 0.14 & 0.17 \\ 
  9 & 5--16, 17--49& 321$\pm$14 & 0.24 & 0.11 & 0.17 \\
  \hline
  \end{tabular}

  $^{\rm a}$ Number of the observation following the notation of
  Table~\ref{tab1}; $^{\rm b}$ Average over the mean countrates of the 100
  lightcurve segments and its standard deviation ([counts/sec/half]).  
  \end{table}

  \begin{figure}
  \resizebox{\hsize}{!}{\includegraphics{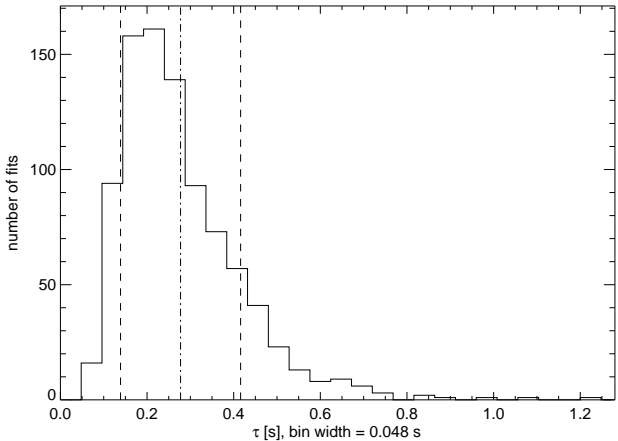}}
  \caption{Distribution of values of the relaxation timescale $\tau$,
  obtained from LSSM[1] fits to a sample of 900 Cyg~X-1\ lightcurves. The
  mean value of the distribution is $\bar\tau=$0.28\,s (dash-dotted line)
  with the standard deviation $s_\tau=$0.14\,s (dashed lines). The most
  probable value is $\tau_{\rm max}=$0.22\,s.}\label{fig5} 
  \end{figure}

  Since the observations provide consistent results, all 900 values for
  $\tau$ can be combined, in order to arrive at a distribution with better
  statistics (Fig.~\ref{fig5}). Note, that this non-Gaussian distribution,
  resulting from the maximum likelihood estimation of $\tau$, cannot be
  described analytically (\cite{hamilton:87}). In addition, simulations
  have shown that the distribution of $\tau$ is contaminated by bad fit
  results at its ``high $\tau$ end'' due to numerical effects
  (\cite{hamilton:87,koenig:97b}). Therefore, $\bar\tau$ can be expected to
  overestimate the ``true'' relaxation timescale of Cyg~X-1, whereas the
  most probable value $\tau_{\rm max}$ should be a better estimator.

\section{Summary and Discussion}\label{kapitel5}

\subsection{The LSSM[1] results --- a generalization
  of the shot noise approach}\label{kapitel5.1} 
  We have shown that the short-term variability of the EXOSAT ME
  lightcurves of Cyg~X-1\ can be well described by a Linear State Space
  Model of first order, i.e.\ the dynamics of the system can be modeled as
  an autoregressive process with one temporal parameter -- the relaxation
  timescale $\tau$. We find $\tau$ to be (0.19$\pm$0.03)\,s for the
  ME energy-range. The relaxator is not found to be correlated either with
  time nor with the luminosity of the source, indicating that the physical
  process producing the emitted radiation is very stable and suggesting
  that the short-term variability is independent of the mass accretion rate
  of the system. The LSSM analysis in the time domain allows us to estimate
  the fraction of the countrate that is due to observation-noise. When
  analyzing the periodogram in the frequency domain this white noise
  component can only be represented by a constant whose extraction is not
  trivial (\cite{belloni:90b,zhangw:95a}). Furthermore, the problem of
  spectral leakage that has to be delt with in the frequency domain, is
  circumvented by fitting in the time domain. For these reasons our LSSM
  analysis delivers results with higher statistical significance than
  corresponding frequency domain fits (\cite{koenig:97a}).

  These LSSM[1] results allow a much simpler description of hard state
  short-term variability than multi-timescale shot noise models. Although a
  detailed quantitative comparison is beyond the scope of this paper, the
  reproduction of the periodogram by the LSSM[1] (compared to its
  approximation by adding shot-profiles with different timescales,
  corrected for observation-noise and binning) as well as the greater
  sensitivity of the LSSM fitting procedure (compared to frequency domain
  fits, which are usually used to evaluate shot noise models) suggest that
  LSSM[1]s are better suited to describe the nature of the observed
  variability.

  The LSSMs can model the different realizations of a stochastic relaxator
  $\tau$, whereas shot noise models are generally restricted by the
  definition of special shot forms. Shot noise lightcurves therefore might
  be regarded as a subclass of LSSM lightcurves in the sense that a
  superposition of exponentially decaying shots can be interpreted as
  \emph{one} possible realization of an intrinsic AR[1] process. The
  inspection of the measured lightcurves of Cyg~X-1 implies that the source
  of the derived AR[1] process is indeed the stochastic superposition of
  individual shot events, corresponding to the basic idea of shot noise. We
  note that on larger timescales this kind of variability is also present
  in the X-ray emission of active galactic nuclei (AGN)
  (e.g.~\cite{mchardy:89,mushotzky:93,koenig:97a,koenig:97c}). The physical
  mechanism responsible for such a temporal behavior, however, is not yet
  understood.

\subsection{The LSSM results in the light of time-dependent Comptonization
  models}\label{kapitel5.2}

  Recently, the discussion concerning accretion physics has begun
  to concentrate on the consideration of timing and spectral properties of
  the X-ray emission as two aspects of the same model
  (\cite{kazanas:97a,koenig:97c,wilms:97c}; and references therein). The
  spectrum of both, AGN and the hard state of galactic black hole
  candidates, is usually explained by inverse Comptonization, where soft
  X-ray photons, provided by a cold accretion disk, are upscattered by
  inverse Compton collisions in a hot plasma to produce the observed high
  energy power-law. 

\begin{figure} \resizebox{\hsize}{!}{\includegraphics{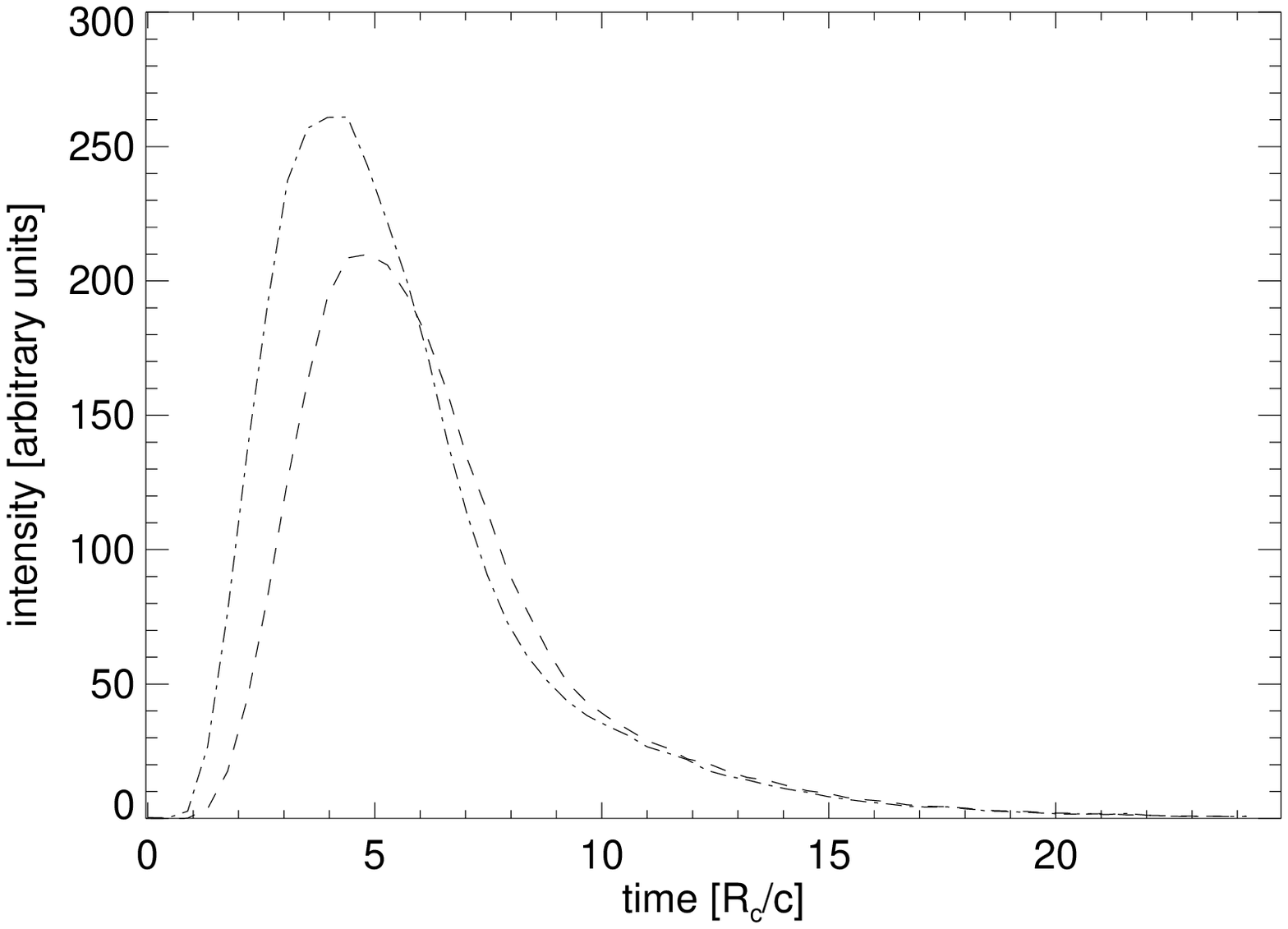}}
  \caption{Shot profiles for the energy-ranges 4--5\,keV (dash-dotted line)
  and 12--23\,keV (dashed line). The profiles have been computed for an
  accretion disk corona (ADC) model with an accretion geometry as described by
  \protect\citey{dove:97b}. This geometry is similar to an advection
  dominated flow (ADAF, \protect\cite{abramo:95,narayan:96}). The ADC model
  parameters were chosen to be appropriate for Cyg~X-1 (disk temperature:
  $kT_{\rm d}$=200\,eV, coronal temperature: $kT_{\rm c}$=66\,keV,
  optical depth of the corona: $\tau_{\rm c}$=2.1;
  \protect\cite{dove:97b}).}\label{fig6} \end{figure}

  In this context it can be assumed that the observed X-ray
  temporal behavior is the result of the processing of short shots of seed
  photons within the accretion disk corona (\cite{payne:80}). The initial
  photon pulse is hardened and temporally broadened while diffusing through
  the corona. As harder photons in the emerging spectrum have, on average,
  undergone more scattering events, they have stayed in the corona for a
  longer time than softer photons and the emerging pulse is comparatively
  broader (observed in AGN by \cite{koenig:97c}). In addition, variability
  structures in high energy lightcurves exhibit a characteristic time-lag
  with respect to those in low energy lightcurves
  (Fig.~\ref{fig6}). Frequency-dependent time-lags have e.g.\ been observed
  in {\sl Ginga} and RXTE observations of Cyg~X-1
  (\cite{miyamoto:88,miyamoto:92,wilms:97c}). The question to what extent
  these lags are produced in a hot corona or wether they are intrinsic
  properties of the cold disk is still unresolved
  (\cite{miyamoto:88,miller:95,nowak:96,nowak:98}). \citey{nowak:98},
  however, show that the frequency-dependent time-lags of Cyg~X-1 can be
  qualitatively reproduced using a simple propagation model (but they also
  point out that the longest observed time-lags ($\approx$0.1\,sec) can
  probably not be explained by Comptonization). Another challenge for
  combined spectral and temporal theories is given by modelling the
  coherence of variability structures in different energy bands which is
  observed to be near unity over a large frequency range in Cyg~X-1
  (\cite{vaughan:97,wilms:97c}).

  Additional support for the time-dependent Comptonization model comes from
  the joint spectral and temporal analysis of AGN, which have similar
  properties to those of Cyg~X-1. For a sample of AGN observations we were
  able to show that a linear correlation between the photon index $\Gamma$
  and the LSSM[1] timescale $\tau$ exists, in the sense that harder spectra
  have a longer variability timescale (\cite{koenig:97c}). By comparing the
  observed $\Gamma$-$\tau$ relation with Monte Carlo simulations of
  time-dependent Comptonization models, it is possible to scale the model
  geometry (\cite{koenig:97b}). For the hard-state galactic black hole
  candidates a similar study cannot be performed since not enough objects
  are known. What can be done, however, is to check whether our best-fit
  LSSM models for Cyg~X-1 could, at least roughly, be explained with a
  time-dependent Comptonization model.
 
  We have computed shot profiles for several possible accretion geometries
  using a linear Monte Carlo code. The detailed results of these
  simulations will be discussed elsewhere.  
  Since shot noise can be regarded as a special representation of an AR[1]
  process (see Sect.~\ref{kapitel5.1}), we can use the computed shots to
  generate lightcurves based on these profiles, including an additional
  white noise component. In analogy to our treatment of the distribution of
  $\tau$ seen in the observations, we derive a most probable value for
  $\tau$ from the simulated lightcurve samples, $\tau_{\rm th}$. We find
  $\tau_{\rm th}=(3\ldots 6) R_{\rm c}/c$, with $R_{\rm c}$ being the
  radius of the corona. The profile of a Compton-shot only depends on the
  relative size of the system, parameterized by the light travel time
  through one radius of the spherical corona, $R_c/c$. By identifying
  $\tau_{\rm th}$ with the measured value of $\tau=$0.19\,s, it is
  therefore possible to express the measured $\tau$ in terms of the coronal
  radius. Our simulations give a first estimate of 320--640 Schwarzschild
  radii for the coronal radius (assuming a mass of 10$M_\odot$ for the
  black hole). Other authors estimate the size of the Comptonization cloud
  in the hard state to be $\approx$100 Schwarzschild radii (\cite{esin:97},
  ADAFs) or $\approx$23 Schwarzschild radii (\cite{nowak:98}, from minimal
  time-lags). These values are only in rough agreement and a better
  understanding of the relation between spectral and temporal properties is
  needed to arrive at a consistent picture. We plan a more detailed study
  of the Compton-shot profiles as well as LSSM analyses of RXTE data to
  further constrain the accretion geometry of Cyg~X-1.

\begin{acknowledgements}
We thank J.~Timmer for his
assistance concerning the LSSM analysis method and C.~Gantert for writing
the LSSM code which has been kindly provided from the ``Freiburger Zentrum
f\"ur Datenanalyse und Modellbildung''. We thank M.~Nowak and J.~Dove for
useful discussions as well as S.~Kitamoto, the referee, for helpful
comments. This research has made use of data obtained through the HEASARC
Online Service, provided by NASA-GSFC.
\end{acknowledgements}

\end{document}